\documentclass[twocolumn,showpacs,preprintnumbers,amsmath,amssymb, prb]{revtex4}
\usepackage{graphicx}% Include figure files
\usepackage{dcolumn}% Align table columns on decimal point
\usepackage{bm}% bold math
\begin{document}

\preprint{APS/123-QED}

\title{Signatures of dynamically polarized nuclear spins in all-electrical lateral spin transport devices}

\author{G. Salis}
%\email{gsa@zurich.ibm.com}
\author {A. Fuhrer}
\author {S. F. Alvarado}

\affiliation{IBM Research, Zurich Research Laboratory,
S\"aumerstrasse 4, 8803 R\"uschlikon, Switzerland}

\date{July 20, 2006}

\begin{abstract}
The effect of nuclear spins in Fe/GaAs all-electrical spin-injection devices is investigated. At temperatures below 50\,K, strong modifications of the non-local spin signal are found that are characteristic for hyperfine coupling between conduction electrons and dynamically polarized nuclear spins.
The perpendicular component of the nuclear Overhauser field depolarizes electron spins near zero in-plane external magnetic field, and can suppress such dephasing when antialigned with the external field, leading to satellite peaks in a Hanle measurement. The features observed agree well with a Monte Carlo simulation of the spin diffusion equation including hyperfine interaction, and are used to study the nuclear spin dynamics and to quantify the Overhauser field that is related to the spin polarization of the injected electrons.
\end{abstract}

\maketitle

The interdependence of nuclear and electron spin dynamics in semiconductors caused by the contact hyperfine interaction leads to a rich variety of phenomena that significantly alter the behavior of independent electron and nuclear systems. For instance, fully polarized nuclear spins in GaAs create an effective magnetic field of 5.3\,T acting on the spin of conductance-band electrons. This interaction has implications for applications in quantum information processing and spintronics. On the one hand it can be employed as a means to efficiently control the electron spin state,~\cite{Reilly2008} but on the other hand it also leads to spin dephasing.
Conduction-band electron spins in a semiconductor can be efficiently polarized either by means of optical orientation~\cite{OpticalOrientation} or spin injection from ferromagnetic contacts.~\cite{Ohno1999,Fiederling1999,Zhu2001,Hanbicki2002}
The hyperfine interaction leads to a flip-flop spin scattering between the electron and nuclear spins that dynamically transfers the spin polarization to the nuclear system.~\cite{Lampel1968, Strand2003} The static part of the hyperfine interaction can be described by an effective (Overhauser) magnetic field $\mathbf B_n$ that acts on the electron spins and has been detected optically~\cite{OpticalOrientation, Paget1977} and in transport experiments.~\cite{Tarucha2004,Petta2008} The restriction of electron spin pumping to small quantum-confined regions allows one to study nuclear polarization in semiconductor heterostructures~\cite{Barrett1994, Marohn1995, Salis2001b} and quantum dots.~\cite{Gammon2001, Tarucha2004, Petta2008} All-electrical injection and detection of electron spins were recently demonstrated in bulk GaAs~\cite{Lou2007, Ciorga2009} and Si,~\cite{Erve2007,Appelbaum2007} and it was suggested that the linewidth of Hanle peaks is influenced by dynamic nuclear polarization (DNP) at lower temperatures.~\cite{Awo-Affouda2009}

\begin{figure}[ht]
\includegraphics[width=80mm]{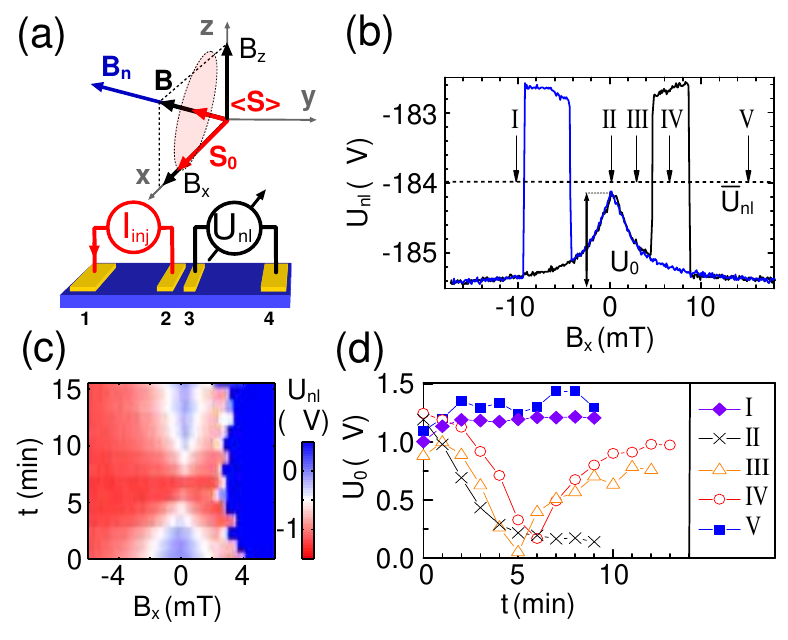}
\caption{\label{fig:fig1} (color online) (a) Coordinate system with external magnetic field and spin vectors and a scheme of the non-local sample geometry and measurement setup with two central ferromagnetic injection (2) and detection(3) bars and two outer reference bars (1) and (4). (b) The nonlocal voltage $U_{\textrm{nl}}$ exhibits jumps for both upward (black) and downward (blue) sweeps of $B_x$. The jumps are related to magnetization switching of the injection and detection bars into parallel and antiparallel configurations. The central peak in $U_{\textrm{nl}}$ is a measure of nuclear polarization as is indicated in a sweep of $B_x$ (c) after waiting for 10 min at $B_x=-50$\,mT and a time $t$ at $B_x^{\textrm{set}}=6.5$\,mT [position IV in (b)]. The fitted height $\Delta U_0$ of the peak vs. $t$ is shown in (d) for different waiting fields $B_x^{\textrm{set}}$ labeled I-V in (b) and (d).}
\end{figure}

Here, we investigate the consequences of DNP in an all-electrical non-local spin device consisting of Fe injection and detection contacts and a $n$-doped GaAs spin transport channel. From measurements of the non-local voltage $U_{\textrm{nl}}$ at the detection contact, we obtain quantitative information on the Overhauser field $B_n$ and on the nuclear spin dynamics in the GaAs channel. By applying an external magnetic field $\mathbf B = B_x \mathbf{\hat{x}} + B_z \mathbf{\hat{z}}$ [see definition of coordinate system and sample layout in Fig.~\ref{fig:fig1}(a)], we investigate the interdependence of nuclear and electron spin dynamics by (i) in-plane sweeps of $B_x$ at constant $B_z$, and by (ii) perpendicular (Hanle) sweeps of $B_z$ at constant $B_x$. For (i), we probe the depolarization peak in $U_{\textrm{nl}}$ at $B_x=0$, previously reported in Ref.~\onlinecite{Lou2007}, which we here explain in terms of a Hanle-type electron spin dephasing arising from a perpendicular $B_n$. In (ii), we observe that apart from the Hanle peak at $B_z=0$, two satellite peaks occur at finite and opposite $B_z$ values provided a finite $B_x$ is oriented parallel to the spin-polarization vector of the injected electrons. We show that these satellite peaks occur when $\mathbf B_n+\mathbf B=0$, leading to a reduction of spin dephasing. A comparison of the measurements with a numerical model allows us to extract quantitative values for $B_n$, the sign of injected electron spins, and a lower limit for injected spin polarization. We find that majority spins are injected into GaAs, and that minority spins get accumulated in GaAs when electrons are extracted from the semiconductor. A lower limit of 1\% for the spin-polarization in the GaAs channel is estimated at 25\,K and a current of 30\,$\mu$A through a contact area of 360\,$\mu$m$^2$.

The spin-devices were prepared by epitaxially growing a 1\,$\mu$m thick n-doped GaAs epilayer with Si doping concentration of $5\times10^{16}$\,cm$^{-3}$ onto an undoped GaAs(001) wafer. The doping concentration within 15\,nm below the surface is $6\times10^{18}$\,cm$^{-3}$, followed by a gradual reduction to $5\times10^{16}$\,cm$^{-3}$ within 15\,nm. The highly-doped surface region allows one to obtain a thin Schottky barrier for efficient charge carrier injection.~\cite{Zhu2001,Hanbicki2002} The substrates, protected by an As capping layer, are then transferred into an ultra-high vacuum chamber for Fe growth by thermal sublimation. Prior to deposition of a 4--6\,nm thick Fe film, the As capping was removed by heating the wafer to 400$^{\circ}$C for one hour. The GaAs surface was inspected by scanning tunneling microscopy to have a c$4\times4$ reconstruction. A final 2--4\,nm thick Au layer protects the Fe film from oxidation. Samples were annealed \emph{in situ} at 220\,$^{\circ}$C for 10 min before further processing. By means of optical or e-beam lithography and ion milling, the Fe layer was patterned into stripes that are 60\,$\mu$m long and 6 and 2\,$\mu$m wide (parallel to the [110] orientation of the GaAs substrate) serving as injection (2) and detection (3) contacts, as indicated in Fig.~\ref{fig:fig1}(a), which shows a scheme of a sample. Unless stated otherwise, the separation between injection and detection contacts was 3\,$\mu$m. A 100\,nm thick layer of Al$_2$O$_3$ isolates large Au/Ti bond pads for contacting the Fe bars from the substrate. Injection and detection of electron spins are achieved in the non-local geometry.~\cite{Jedema2001,Lou2007} A current $I_{\textrm{inj}}$ is drawn from contact (1) to contact (2) such that spin is injected at contact (2) for $I_{\textrm{inj}}>0$ and spin filtering occurs for $I_{\textrm{inj}}<0$. The nonlocal voltage $U_{\textrm{nl}}$ is measured between contact (3) and contact (4) using both dc and ac lock-in techniques. Both approaches yield equivalent results, and in the following we use a superscript to differentiate ac ($I_{\textrm{inj}}^{ac}$) from dc ($I_{\textrm{inj}}$) excitation of the injection current. Measurements were performed in two different cryostats with variable-temperature inserts and a superconducting magnet system. One of them allows application of magnetic fields $B_x$ and $B_z$ along two independent axes.
%Both systems also permit in-situ sample rotation.

Figure~\ref{fig:fig1}(b) shows the nonlocal voltage $U_{\textrm{nl}}$ measured as a function of $B_x$ at temperature $T=5$\,K, obtained at $I_{\textrm{inj}}^{ac}$=1.6\,$\mu$A. In an upward sweep of $B_x$, $U_{\textrm{nl}}$ jumps to a higher value at $B_x\approx5$\,mT when the first bar reverses its magnetization, and drops back down at the reversal of the second bar, i.e., when the magnetizations are parallel again. In the following, we subtract an offset $\bar{U}_{\textrm{nl}}$ from $U_{\textrm{nl}}$ so that $U_{\textrm{nl}}=0$ lies exactly in the middle between the two jumps, marking the nonlocal voltage level with zero electron spin polarization. In addition to the jumps, $U_{\textrm{nl}}$ forms a peak at $B_x=0$\,mT, indicative of a loss of average electron spin polarization at contact (3). Its height $\Delta U_0$ depends on the history before performing the $B_x$ sweep and appears within a time scale of several minutes, which is characteristic of nuclear spin-lattice relaxation times $\tau_1$.~\cite{Lu2006}

To demonstrate that the depolarization peak is related to nuclear spin polarization, we performed a series of measurements in which DNP was built up and then reversed while monitoring $\Delta U_0$. The system is initialized at $B_x=-50$\,mT for 10 min with $I_{\textrm{inj}}=-1.6$\,$\mu$A such that nuclear spins get dynamically polarized until they reach a saturation value. Then $B_x$ is swept to a value $B_x^{\textrm{set}}$. The nuclear spins adiabatically follow the external field, and if $B_x$ crosses zero reverse their direction in space.~\cite{VanDorpe2005} After waiting a time $t$ at $B_x=B_x^{set}$, the depolarization peak is immediately recorded by sweeping $B_x$ across $B_x=0$\,mT, with $I_{\textrm{inj}}^{ac}=1.6$\,$\mu$A. This was repeated for several values of $t$ to obtain data as shown in Fig.~\ref{fig:fig1}(c) for $B_x^{\textrm{set}}=6.5$\,mT [corresponding to arrow IV in Fig.~\ref{fig:fig1}(b)], where $\Delta U_0$ first decreases, passes through a minimum at $t\approx$\,6\,min and saturates again at $t>$\,10\,min. This time-dependence of $\Delta U_0$ is a strong indication that it is a measure of the nuclear polarization. Substantial nuclear spin polarization $\langle\mathbf{I}\rangle$ can be built up by hyperfine-induced flip-flop spin scattering if an average electron spin polarization $\langle\mathbf{S}\rangle$ is sustained, such as in our case by injection or filtering of spin-polarized electrons from the Fe contact (2). The effective Overhauser magnetic field $\mathbf{B}_n\propto \langle\mathbf{I}\rangle$, which in steady state can be described by~\cite{OpticalOrientation}

\begin{equation}\label{eq1}
\mathbf{B}_n=f b_n {\frac{\mathbf{B}\cdot\langle\mathbf{S}\rangle}{{B^2}}}\mathbf{B}.
\end{equation}

Here, $f\leq1$ is a leakage factor that takes into account the possibility of nuclear spin relaxation by other channels than through a hyperfine-induced flip-flop process, and $b_n=-5.3$\,T in GaAs~\cite{Paget1977} is the maximum field for 100\% nuclear spin polarization. Equation~(\ref{eq1}) neglects the Knight shift and the effect of dipole-dipole interaction between nuclear spins that is only important for typically $B<1$\,mT, where it leads to a drop of $\langle \mathbf{I} \rangle$ to zero at $B=0$. In Eq.~\ref{eq1}, $\langle \mathbf{S} \rangle$ can be replaced by $\mathbf{S}_0=S_0 \mathbf{\hat{x}}$, which denotes the average spin polarization of the GaAs electron density without precession, see Fig.~\ref{fig:fig1}(a). Depending on the sign of $B_x S_0$, $\mathbf{B}_n$ is aligned parallel or antiparallel to $\mathbf{B}$. When $B_x$ changes sign, the spatial direction of DNP does not change, which means that after adiabatic reversal of the nuclear polarization at $B_x=0$\,mT, the nuclear polarization will first decrease and then repolarize into the opposite direction. This is exactly what is observed in $\Delta U_0$. A similar decrease and subsequent increase of $\Delta U_{0}$ are measured if the sign of $I_{\textrm{inj}}$ is reversed, whereby electron spins with the opposite sign will accumulate below the injection contact (data not shown). In Fig.~\ref{fig:fig1}(d), $\Delta U_{0}$ is plotted as a function of $t$ for different $B_x^{\textrm{set}}$, labeled I to V in Fig.~\ref{fig:fig1}(b). For position (I) and (V), $\Delta U_{0}$ does not fall to zero but slightly increases before saturation. Because $B_x^{\textrm{set}}$ remains negative (I) or is positive and large enough such that the magnetization of both injection and detection contact reverses (V), no crossing of $\langle \mathbf{I} \rangle=0$ is necessary to attain steady state. Only for the two intermediate fields (III and IV), where $B_x$ has crossed zero but the magnetization of the injection contact has not yet reversed, will $\Delta U_{0}$ drop to zero and reappear afterwards. For $B_x^{\textrm{set}}=0$, build-up of nuclear spin polarization is prevented, see trace II in Fig.~\ref{fig:fig1}(d), because of inefficient DNP for $\mathbf B \cdot \langle \mathbf{S} \rangle=0$.

\begin{figure}[ht]
\includegraphics[width=80mm]{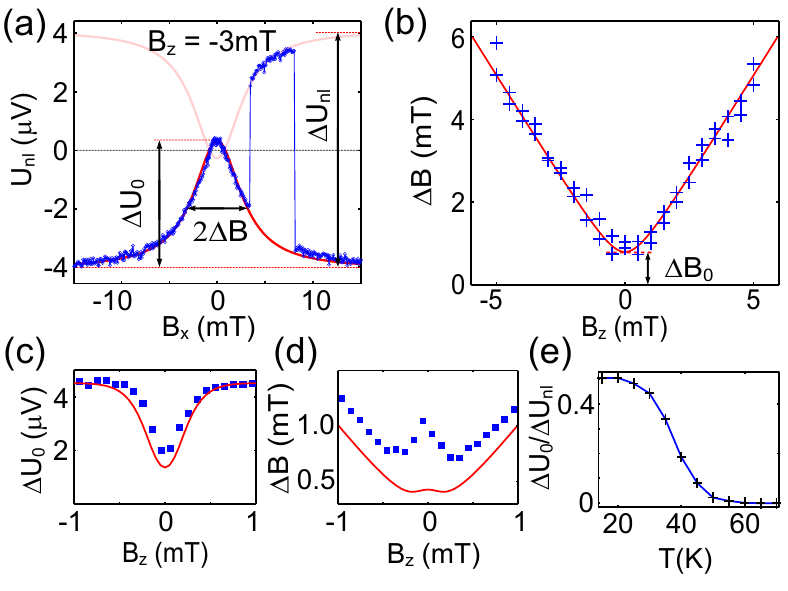}
\caption{\label{fig:fig2} (color online) (a) $U_{\textrm{nl}}$ as a function of $B_x$ at a fixed $B_z=-3$\,mT at 20\,K. The red curve is a Lorentzian with a half width $\Delta B=-3$\,mT and a height $\Delta U_0=4.2$\,$\mu$V that has been adjusted for best fit to the depolarization peak at $B_x=0$. For antiparallel magnetization of the central contacts the Lorentzian is inverted and offset by the maximum spin signal $\Delta U_{\mathrm nl}$. (b) Peak width $\Delta B$ as a function of $B_z$. At vanishing $B_z$, $\Delta B$ saturates at a finite value $\Delta B_0=0.8$\,mT. (c) $\Delta U_0$ (squares) dips at $B_z=0$\,mT because of the nuclear local dipolar field. The solid red line is calculated using a model as described in the text. (d) Measured values of $\Delta B$ around $B_z=0$\,mT (squares), showing a local peak, and model (solid red line) as described in the text. (e) Temperature dependence of $\Delta U_0/\Delta U_{\textrm{nl}}$, which is slightly larger than $0.5$ for low temperatures and vanishes for $T\geq50$\,K.}
\end{figure}

To understand the reason for the occurrence of the peak in $U_{\textrm{nl}}$, we have carried out measurements of $U_{\textrm{nl}}$ versus $B_x$ for different $B_z$ shown in Fig.~\ref{fig:fig2}. The nuclear spin polarization is initialized at $I_{\textrm{inj}}=20$\,$\mu$A and $B_x=-50$\,mT for 15 min. $B_x$ is then swept up and down with a sweep rate of 25\,mT/min, for different $B_z$ from $-6$ to 6\,mT in steps of 0.5\,mT. Figure~\ref{fig:fig2}(a) shows a trace measured at $B_z=-3$\,mT and $T=20$\,K. $\Delta U_{0}$ extends slightly beyond $\Delta U_{\textrm{nl}}/2$, defined as half the separation between $U_{\textrm{nl}}$ for parallel and antiparallel magnetization. The peak can be fitted by a Lorentz curve $\propto (1+B_x^2/\Delta B^2)^{-1}$, where the half width at half maximum $\Delta B$ of the peak follows $(B_z^2+\Delta B_{0}^2)^{1/2}$ with $\Delta B_{0}=0.8$\,mT, see Fig.~\ref{fig:fig2}(b). The Lorentz shape with a width $\Delta B=B_z$ suggests that the depolarization peak is due to the rotation of the total magnetic field $\mathbf B_{\textrm{tot}}=\mathbf{B}+\mathbf{B}_n$ in the $xz$ plane as $B_x$ is swept through zero. For sufficiently large $B_{\textrm{tot}}$, the electron spins precess fast enough that $\langle \mathbf S \rangle$ points along (or against) $\mathbf B_{\textrm{tot}}$. $U_{\textrm{nl}}$ is given by the projection of $\langle \mathbf S \rangle$ onto $\mathbf{\hat x}$, and thus becomes proportional to $B_x^2/(B_x^2+B_z^2)$, i.e., follows a Lorentz curve with a half width equal to $B_z$, as observed in the experiment for larger $|B_z|$ [the red curve in Fig.~\ref{fig:fig2}(a) is a Lorentzian fit with $\Delta B=|B_z|=3$\,mT and $\Delta U_0$ as the only free fit parameter].

Next we discuss why $\Delta B$ and $\Delta U_0$ do not disappear at $B_z=0$. We have repeated measurements as the ones shown in Fig.~\ref{fig:fig2}(b), but with higher resolution around $B_z=0$ and at $T=25$\,K. As shown in Fig.~\ref{fig:fig2}(c), a dip in $\Delta U_0$ appears at $B_z=0$\,mT with a full width at half maximum of about 0.5\,mT and a decrease from 4.5 to 2\,$\mu$V. This is evidence of the presence of a small field component $B_y$ along $\mathbf{\hat{y}}$ that orients $\mathbf B_n$ into the $y$ direction, accompanied by a partial depolarization of the nuclear spins because of dipole-dipole interaction between the nuclear spins. For a local dipole field $B_L$, $B_n = B_n^0 B^2/(B^2+B_L^2)$, where $B_n^0$ is the nuclear field for $B \gg B_L$. In GaAs, $B_L\approx 1$\,mT.~\cite{Paget1977} An expression for $U_{\textrm{nl}}$ is obtained for arbitrary $B_{\textrm{tot}}$ by separating $\langle \mathbf S \rangle$ into its components along and perpendicular to $\mathbf B_{\textrm{tot}}$. We find $U_{\textrm{nl}} = -{\Delta U_{\textrm{nl}}\over2} (\cos^2 \alpha + H(B_{\textrm{tot}})/H(0) \sin^2 \alpha )$, where $H(B_{\textrm{tot}})$ is the Hanle lineshape as defined in Eq.~(1) of Ref.~\onlinecite{Lou2007} and $\alpha = \arctan \sqrt{B_y^2+B_z^2}/B_x$. The term proportional to $\cos^2 \alpha$ ($\sin^2 \alpha$) corresponds to the component of $\langle \mathbf S \rangle$ parallel (perpendicular) to $\mathbf B_{\textrm{tot}}$. $\Delta U_0$ is given by the value of $U_{\textrm{nl}}$ at $\alpha=0$, ${\Delta U_{\textrm{nl}}\over2} (1-H(B_{\textrm{tot}})/H(0))$, and therefore follows a typical Hanle curve: For $B<B_L$, $\Delta U_0$ decreases because $B_n$ depolarizes and thus $B_{\textrm{tot}}$ decreases, making Hanle-type spin dephasing less efficient. For intermediate $B_{\textrm{tot}}$, $H(B_{\textrm{tot}})$ becomes negative and therefore $\Delta U_0>\Delta U_{\textrm{nl}}/2$. The solid line in Fig.~\ref{fig:fig2}(c) shows the calculated $\Delta U_0$ using the model described above and reproducing the observed dip. As parameters, we used a diffusion constant $D=0.002$\,m$^2$/s, spin lifetime $\tau_s=10$\,ns, $B_n^0=47$\,mT, $B_L=1$\,mT and $B_y=0.2$\,mT. Interestingly, in this model, $\Delta B$ does not drop to zero, but even increases around $B_z=0$\,mT. Figure~\ref{fig:fig2}(d) shows the measured $\Delta B$ having a local peak at $B_z=0$, as well as the results of a fit to the model using the same parameters as above (solid red line). In the model, the local increase in $\Delta B$ is because of a change of the form of the depolarization peak whose height $\Delta U_0$ decreases while its tails remain unchanged because they are determined by the term proportional to $\cos^2 \alpha$. Compared to the model, the measured $\Delta B$ exhibits a more pronounced local peak at $B_z=0$. We note that the sizes of $\Delta U_0$ and $\Delta B$ at $B_z=0$ are determined by the interplay of many different mechanisms, from which our model takes into account the electron spin dynamics along the diffusive path between injection and detection contacts, a homogeneous Overhauser field aligned with the external field, a decrease of nuclear polarization at $B=0$ because of the local dipole field, and a finite magnetic field component along $\mathbf{\hat{y}}$. We have neglected the Knight shift that reorients the direction of the Overhauser field, as well as the possibility of electron spin dephasing in a locally fluctuating Overhauser field, effects that both will affect the details of $U_{\textrm{nl}}$ around $B=0$.

The temperature dependence of $\Delta U_0/\Delta U_{\textrm{nl}}$ at $B_z \approx 1$\,mT is shown in Fig.~\ref{fig:fig2}(e). For $T<20$\,K, $\Delta U_0/\Delta U_{\textrm{nl}}$ remains slightly above 0.5, indicative of the intermediate field regime. At $T\geq 25$\,K, $\Delta U_0/\Delta U_{\textrm{nl}}$ decreases. Within the explanation given above, $B_\textrm{tot}$ must there be of the same order as the Hanle peak width, i.e. about 3\,mT. From measurements of satellite peaks in Hanle configurations (explained later), we can determine $B_n$ and extrapolate values well above 20\,mT for 25\,K and initialization at $B_x=-50$\,mT, which is much larger than 3\,mT, thus supporting the model that involves the nuclear dipolar mechanism that reduces $B_n$. Above $T=50$\,K, where DNP becomes inefficient, $\Delta U_0$ is no longer observable.

\begin{figure}[ht]
\includegraphics[width=80mm]{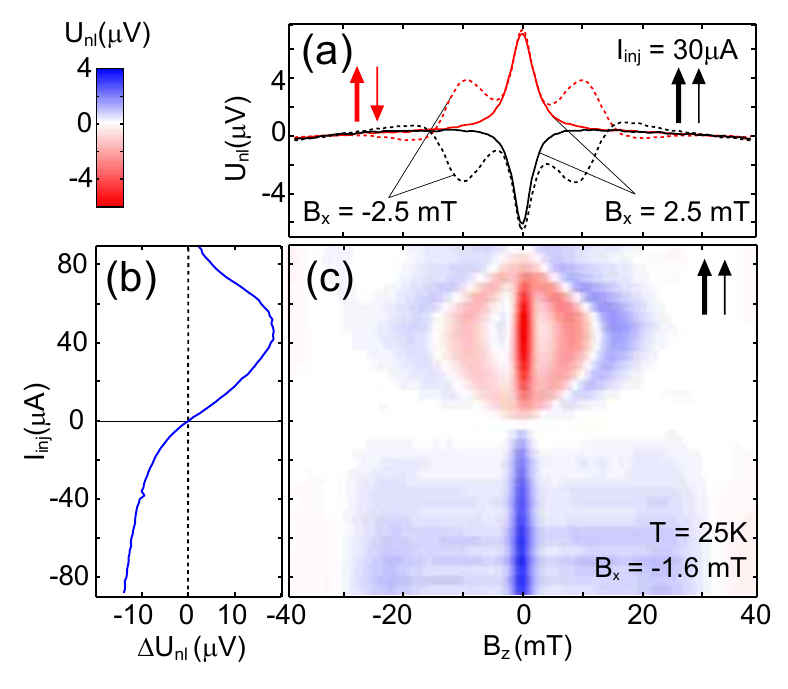}
\caption{\label{fig:fig3} (a) Hanle measurements of $U_{\textrm{nl}}$ versus $B_z$ for fixed $B_x=\pm2.5$\,mT and parallel (antiparallel) magnetization of the central contact bars indicated by black (red) arrows. (b) Maximum spin signal $\Delta U_{\textrm{nl}}$ as a function of $I_{\textrm{inj}}$. For spin injection, i.e.  $I_{\textrm{inj}}>0$, the spin signal goes through a maximum with increasing $I_{\textrm{inj}}$ that correlates with the satelite peak position in the Hanle measurement in (c) as a function of $B_z$ and $I_{\textrm{inj}}$. For spin filtering, i.e. $I_{\textrm{inj}}<0$, no satellite peaks appear because the nuclear field $B_n$ is parallel to the externally applied field. Data in (a) and (c) were averaged over an up and down sweep of $B_z$.}
\end{figure}

\begin{figure}[ht]
\includegraphics[width=80mm]{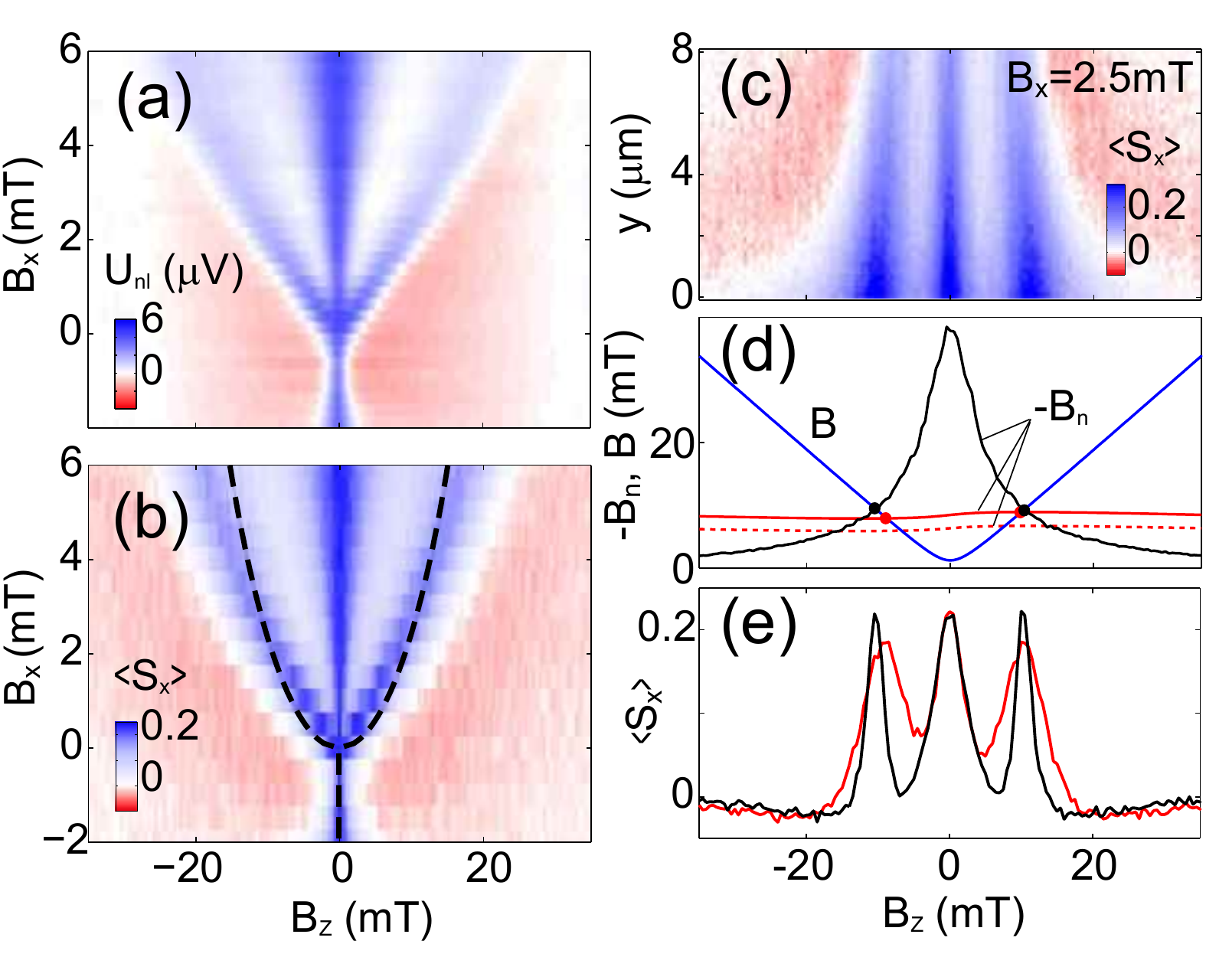}
\caption{\label{fig:fig4} (a) $U_{\textrm{nl}}$ versus $B_z$ for different $B_x$, measured at $I_{\textrm{inj}}=-30$\,$\mu$A and $T=25$\,K. Magnetizations were set parallel along the positive $x$ direction. (b) Calculated spin polarization $\langle S_x \rangle$ at detection contact using Monte Carlo simulation. $D=0.002$\,m$^2$s$^{-1}$, $\tau_s=10$\,ns , $b_n f S_0=45$\,mT in (b)-(e). The dashed black lines indicate expected peak positions for a saturated $B_n$ according to Eq.~\ref{eq1}. (c) Map of $\langle S_x \rangle$ versus $y$ and $B_z$ for $B_x=2.5$\,mT. Spin is injected at $y<0$, and data in (b) is averaged at the detection contact located at $3$\,$<y<5$\,$\mu$m. (d) $B$ (blue) and $-B_n$ versus $B_z$ for $B_x=2.5$\,mT. $B_n$ was calculated for saturated condition (black) and for $r\widetilde{\tau}_1=250$\,mT (red) averaged at the detection contact (dashed line) and in between the contacts (solid lines). Dots indicate positions where $B=-B_n$ and satellite peaks are expected. (e) Hysteresis and broadening of satellite peaks in calculated $\langle S_x \rangle$ at detection contact for locally varying $B_n$ [red, same parameters as in (d)] and uniform/saturated $B_n$.}
\end{figure}

As mentioned in Ref.~\onlinecite{Awo-Affouda2009}, nuclear spin polarization also modifies the lineshape of the Hanle curve $U_{\textrm{nl}}$ versus $B_z$. An even more profound effect occurs when $B_n$ points against $B$ and is so large that the two fields cancel. In such a situation, Hanle-type spin dephasing is strongly reduced, leading to two satellite peaks in $U_{\textrm{nl}}$ at finite and opposite $B_z$ values. Figure~\ref{fig:fig3}(a) shows such measurements at fixed $B_x=2.5$\,mT  (solid lines) and $-2.5$\,mT (dashed lines), and for the magnetization of the detection contact oriented parallel (black) and antiparallel (red) to that of the injection contact. The magnetization of the latter is oriented along positive $B_x$. According to Eq.~(\ref{eq1}), $B_n$ points against or along $B$, depending on the sign of $B_x S_0$. The appearance of the satellite peaks requires that $B_x S_0>0$. For $I_{\textrm{inj}}>0$, we observe the satellite peaks at $B_x<0$ and for $I_{\textrm{inj}}<0$ at $B_x>0$ [see Fig.~\ref{fig:fig4}(a)]. Therefore, $S_0$ is negative (and antialigned with the magnetization $\mathbf{M}$ of the injection contact) in the case of spin injection and positive (aligned with $\mathbf{M}$) for spin filtering. In agreement with previous observations,~\cite{Crooker2005b,Lou2007} this means that majority spins are injected from Fe into GaAs.

The position of the satellite peaks provides a direct measure of the nuclear field because there, $\mathbf{B}=-\mathbf{B_n}$. The sign and magnitude of $S_0$ can be controlled with the injection current $I_{\textrm{inj}}$. As shown in Fig.~\ref{fig:fig3}(b), $\Delta U_{\textrm{nl}}$
reverses its sign at $I_{\textrm{inj}}=0$. For $I_{\textrm{inj}}>0$, $\Delta U_{\textrm{nl}}$ reaches a peak and decreases again, whereas for negative $I_{\textrm{inj}}$ it saturates. The separation of the satellite Hanle peaks measured at $B_x=-1.6$\,mT and shown in Fig.~\ref{fig:fig3}(c) follows the same behavior as $\Delta U_{\textrm{nl}}$, indicating that the nuclear field monotonically depends on $S_0$. This is also evidence that the peak in $\Delta U_{\textrm{nl}}$ for $I_{\textrm{inj}}>0$ directly reflects a maximum spin polarization in the GaAs channel and is not due to a dependence of the detector sensitivity on the injection current that could indirectly occur through a spreading resistance. Because of the opposite direction of $\mathbf S_0$ for $I_{\textrm{inj}}<0$, no satellites are observed in Fig.~\ref{fig:fig3}(c). Similarly, the appearance of the satellites can be controlled by the orientation of the magnetization of the injection contact (data not shown).

Figure~\ref{fig:fig4}(a) shows a colorscale plot of measured $U_{\textrm{nl}}$ as a function of $B_x$ and $B_z$. For small $B_x$, we observe a linear dependence of the satellite peak separation on $B_x$. From the condition $\mathbf B + \mathbf B_n=0$ and Eq.~\ref{eq1}, i.e. when assuming that $B_n$ is at its saturation value for all measured $B$, the satellite peak positions are given by $B_z=\pm \sqrt{B_x ( - f b_n S_0 - B_x)}$. As $-f b_n S_0\gg B_x$ for our measurements, we expect a quadratic dependence, $B_x \propto B_z^2$, at the satellite peaks. In an optical orientation measurement with oblique magnetic field, similar satellite peaks were observed in the circular polarization of photoluminescence as a function of $B_z$.~\cite{Farah1998} Also there, a linear increase of the satellite peak separation was observed, which was interpreted as a $B$-dependent leakage factor $f$. As we will demonstrate with a numerical simulation, our data can be explained without assuming a field-dependent $f$, but taking into account the long $\tau_1$ of nuclear spins whose polarization does not reach saturation at individual field values within a sweep of $B_z$. We performed the numerical simulation of the diffusing electron spins using a Monte Carlo approach by assigning one-dimensional spatial coordinates $y_i$, velocities $v_i$ and three-dimensional spin directions $\mathbf s_i$ to electrons labeled $i=1$ to $n$. At constant time intervals $\delta t$, $y_i$ are updated to $y_i + v_i \delta t$, and to a fraction of the $n$ electrons a new random velocity is assigned, thus simulating the diffusive scattering process characterized by the diffusion constant $D$. The new velocity $v_i$ is distributed between $-v_F$ and $v_F$ according to the projection of a two-dimensional vector of length $v_F$ onto the $y$ axis. Spin coordinates $\mathbf s_i$ are regularly updated by calculating the rotation about the locally varying $\mathbf B+\mathbf B_n$ and by accounting for a spin decay at rate $1/\tau_s$. At a constant rate, spin-polarized electrons are injected by assigning coordinates $y_i$ within the injection contact area ($-6$ to 0 \,$\mu$m) to new electrons $i$. We let $B_n$ locally evolve with a time constant $\tau_1$ towards the saturation nuclear field as calculated by Eq.~\ref{eq1}, thus accounting for the fact that typical sweep rates $r$ in the experiment are faster than $1/\tau_1$. We neglect nuclear spin diffusion because of the small diffusion constant (10$^3$\,{\AA}/s has been measured in Ref~\cite{Paget1982}). The simulation is run for a time $5 \tau_s$, ensuring a converging self-consistent solution.

The nonlocal voltage $U_{\textrm{nl}}$ is proportional to the electron spin component $\langle S_x \rangle$ averaged over the detector contact at 3\,$\mu$m\,$<y<$\,5\,$\mu$m, and is plotted in Figure~\ref{fig:fig4}(b) as a function of $B_x$ and $B_z$. In Fig.~\ref{fig:fig4}(c), a map of $\langle S_x \rangle$ versus $y$ and $B_z$ is shown for $B_x=2.5$\,mT. In the simulation, $\tau_s$\,=\,10\,ns, $D=2 \times 10^{-3}$\,m$^2$/s and $r \tau_1=250$\,mT are used. We obtain an excellent match with the experimental data in Fig.~\ref{fig:fig4}(a) with $f b_n S_0=45$\,mT, where $S_0$ is the averaged $x$-component of $\langle \mathbf S \rangle$ for $0<y<3$\,$\mu$m and at $B_z=0$. The solid line in Fig~\ref{fig:fig4}(b) indicates the increase with $\sqrt{B_x}$ of the satellite peak separation that is expected when $B_n$ reaches its saturated value for all field positions. In contrast to this, the simulation reproduces the linear increase for small $B_x$.
In Fig.~\ref{fig:fig4}(d), calculated $-B_n$ (red) is shown versus $B_z$ averaged in between the two contacts (solid line) and below the detection contact (dashed line). Because $r\tau_1$ is much larger than the sweep range of $\pm40$\,mT, $B_n$ does not follow the saturated value as $B_z$ is swept (shown as black line), but is rather uniform at $B_n\approx - f b_n S_0 B_x \langle {1\over B} \rangle$, where $\langle {1\over B} \rangle$ is the time-average of ${1\over B}$ for a sweep of $B_z$, and $S_0$ is averaged in between the contacts. From this, a splitting that is linear in $B_x$ directly follows. In addition, $B_n$ exhibits a small asymmetry with $\pm B_z$, leading to an asymmetry of the two satellite peak positions, as shown in Fig.~\ref{fig:fig4}(e).
From the data in Fig.~\ref{fig:fig4}(d) one sees that $B_n$ depends on $y$. Therefore $\mathbf B=-\mathbf B_n$ is not fulfilled over the entire distance between injection and detection contacts, leading to a reduced height of the satellite peaks. In Fig.~\ref{fig:fig4}(e), the red line is a linecut through the data in Fig.~\ref{fig:fig4}(b) at $B_x=2.5$\,mT, whereas for the data of the black line, $B_n$ is uniformly fixed to the saturation value predicted by Eq.~\ref{eq1} with $f b_n S_0=45$\,mT. In the latter case, the satellite peaks reach the full height because at $B=-B_n$, the total field disappears everywhere in the sample. The decrease of the satellite peak height is significantly  underestimated in the simulation, compare with Fig.~\ref{fig:fig3}(a). This indicates that $B_n$ might even be more inhomogeneous in the sample than estimated with the Monte Carlo simulation.

The measured size of $f b_n S_0$=45\,mT allows a lower estimate of the injected spin polarization. The value $b_n=-5.3$\,T known from literature (Ref.~\onlinecite{Paget1977}) limits $S_0$ to about 1\% for $f=1$. To obtain a rough estimate of the polarization of the injected current from this, we have to account for the ratio of injected electrons to the $5\times 10^{16}$\,cm$^{-3}$ electrons that are already in the sample. Within a spin lifetime $\tau_s$, $I_{\textrm{inj}}\tau_s/e=1.3\times10^{6}$ electrons are injected and diffuse into a volume 60\,$\mu$m\,$\times($6\,$\mu$m$+\sqrt{(D\tau_s)})\times1$\,$\mu$m, corresponding to a density of 2.1$\times$10$^{15}$\,cm$^{-3}$, i.e. the injected spins make up about $1/24$ of the electron density. Accordingly, the spin polarization of injected electrons is at least 20\% for $f=1$, $I_{\textrm{inj}}=-30$\,$\mu$A and $T=25$\,K.

In conclusion, we have found that the non-local voltage $U_{\textrm{nl}}$ in an all-electrical spin injection and detection device exhibits distinct signatures of dynamically polarized nuclear spins that can be used to measure the Overhauser effective magnetic field $B_n$ and to study nuclear spin dynamics. We obtained a quantitative understanding of the depolarization peak in an in-plane magnetic field sweep. Because the peak height sensitively depends on small stray fields on the order of 0.1\,mT and because of nuclear dipole-dipole interaction, a quantitative relation between the shape/size of the peak and $B_n$ is difficult to obtain. However, a quantitative measurement of $B_n$ is achieved by observing the satellite peaks that occur in a Hanle measurement when $\mathbf B+\mathbf B_n=0$. By comparison with a self-consistent simulation of spin diffusion and hyperfine interaction, we obtain a value for $b_n f \langle S_x \rangle$ of 45\,mT at 25\,K and $I_{\textrm{inj}}=-30$\,$\mu$A, from which the sign of injected spin polarization can be determined and its magnitude estimated. We can explain our data using a leakage factor $f$ that does not depend on the external magnetic field. The observed nuclear spin signatures enable the study of nuclear spin dynamics including nuclear spin resonance in small semiconductor/ferromagnet structures by a transport measurement. Of specific interest is to extend this method to investigate hyperfine interaction in other semiconductor materials like silicon~\cite{Appelbaum2007} or graphene~\cite{Tombros2007} where the spins can not easily be accessed optically.

We acknowledge fruitful discussions with Rolf Allenspach, Reto Schlittler and Leo Gross, and technical support from Meinrad Tschudy, Daniele Caimi, Ute Drechsler and Martin Witzig.

\end{document}